\pdfoutput=1
\documentclass[11pt]{article}
\usepackage{jheppub}
\usepackage[all]{xy}
\usepackage{customprelude}

\usepackage{tabu}
\usepackage{multirow}

\newcommand\be{\begin{equation}}
\newcommand\ee{\end{equation}}

\title{Global Structures from the Infrared}

\author[\dagger]{Michele Del Zotto} 
\author[\sharp]{and Iñaki García Etxebarria}

\affiliation[\sharp]{Department of Mathematical Sciences\\
Durham University, Durham, DH1 3LE, United Kingdom}
\affiliation[\dagger]{Department of Mathematics and Department of Physics and Astronomy\\ Uppsala University, Uppsala, Sweden}

\emailAdd{michele.delzotto@math.uu.se}
\emailAdd{inaki.garcia-etxebarria@durham.ac.uk}

\abstract{
  Quantum field theories with identical local dynamics can admit
  different choices of global structure, leading to different
  partition functions and spectra of extended operators. 
  Such choices can be
  reformulated in terms of a topological field theory in one
    dimension higher, the symmetry TFT.  In this paper we show that
  this TFT can be reconstructed from a careful analysis of the
  infrared Coulomb-like phases. In particular, the TFT matches between the UV and the IR. This provides a purely field
  theoretical counterpart of several recent results obtained via
  geometric engineering in various string/M/F theory setups for
  theories in four and five dimensions that we confirm and extend.}

\begin{document}


\maketitle

\section{Introduction}

Among the most important questions about the dynamics of quantum
fields is the task of characterising features that are robust under
renormalization flow. One such feature is the anomalous behaviour of
global symmetries \cite{tHooft:1979rat},\footnote{\ In this paper by anomalies we always
  mean 't Hooft anomalies for global symmetries.} which are often
captured via inflow \cite{CALLAN1985427} from an invertible field theory in one dimension
higher, known as the \emph{anomaly theory} \cite{Freed:2014iua}.

\medskip

This inflow picture can be enriched whenever the QFT has local
dynamics compatible with inequivalent global structures (these can be
detected by analysing multiple related properties of the theory: the
spectra of non-local operators, the spectrum of generalised
symmetries, or the partition functions on compact curved
spacetimes). When that happens we can split off the choice of global
form from the behaviour of the local degrees of freedom, by viewing
the local degrees of freedom of the QFT as a theory relative
\cite{Freed:2012bs} to a non-invertible theory in one dimension
higher. A proper QFT, with a fully specified global form, can then be
interpreted as the compactification of the bulk non-invertible theory
on an interval: on one end of the interval the bulk theory couples to
a gapped TFT responsible for choosing the global structure, and on the
other end of the interval the bulk theory couples to the local degrees
of freedom.\footnote{\ It is not always the case that diffeomorphism
  invariant choices of gapped boundary conditions exist. The 6d
  $(2,0)$ theories associated to the algebra $\fg$ are examples of
  consistent local dynamics whose associated bulk theories are
  expected not to admit such choices for generic $\fg$. (See
  \cite{Garcia-Etxebarria:2019cnb,Gukov:2020btk} for an analysis of
  which $\fg$ are expected to admit diffeomorphism invariant gapped
  boundary conditions.) This is one way to understand statements in
  the literature that such theories are not ``genuine QFTs''. We
  expect that the bulk theories associated to the 4d $\cN=2$ theories
  studied in this paper always admit such choices.}  This structure is
well-known in the context of Lagrangian theories --- for an
enlightening discussion about the $SU(N)$ versus $PSU(N)$ cases we
refer our readers to section 6 of \cite{Gaiotto:2014kfa}.

\medskip

It is generically the case that the resulting QFT after reduction on
the interface is anomalous, in which case the picture above needs to
be further refined, as elaborated on in \cite{Apruzzi:2021nmk} (see
also \cite{Gukov:2020btk}): we consider a non-invertible theory on an
interval, where a gapped interface on the left connects it to the
anomaly theory for the QFT of interest (different global forms have
different symmetries and therefore different anomaly theories, which
are connected via suitable interfaces to the same non-invertible
theory), and the gapless boundary on the right encodes the local
degrees of freedom of the QFT. So in this case it is the gapped
interface that encodes the choice of global form. As in
\cite{Apruzzi:2021nmk}, we refer to the non-invertible theory inside
the interval as the \emph{symmetry theory}.

\medskip

\begin{figure}
    \centering
    \begin{tabular}{cc}
    \includegraphics[scale=0.35]{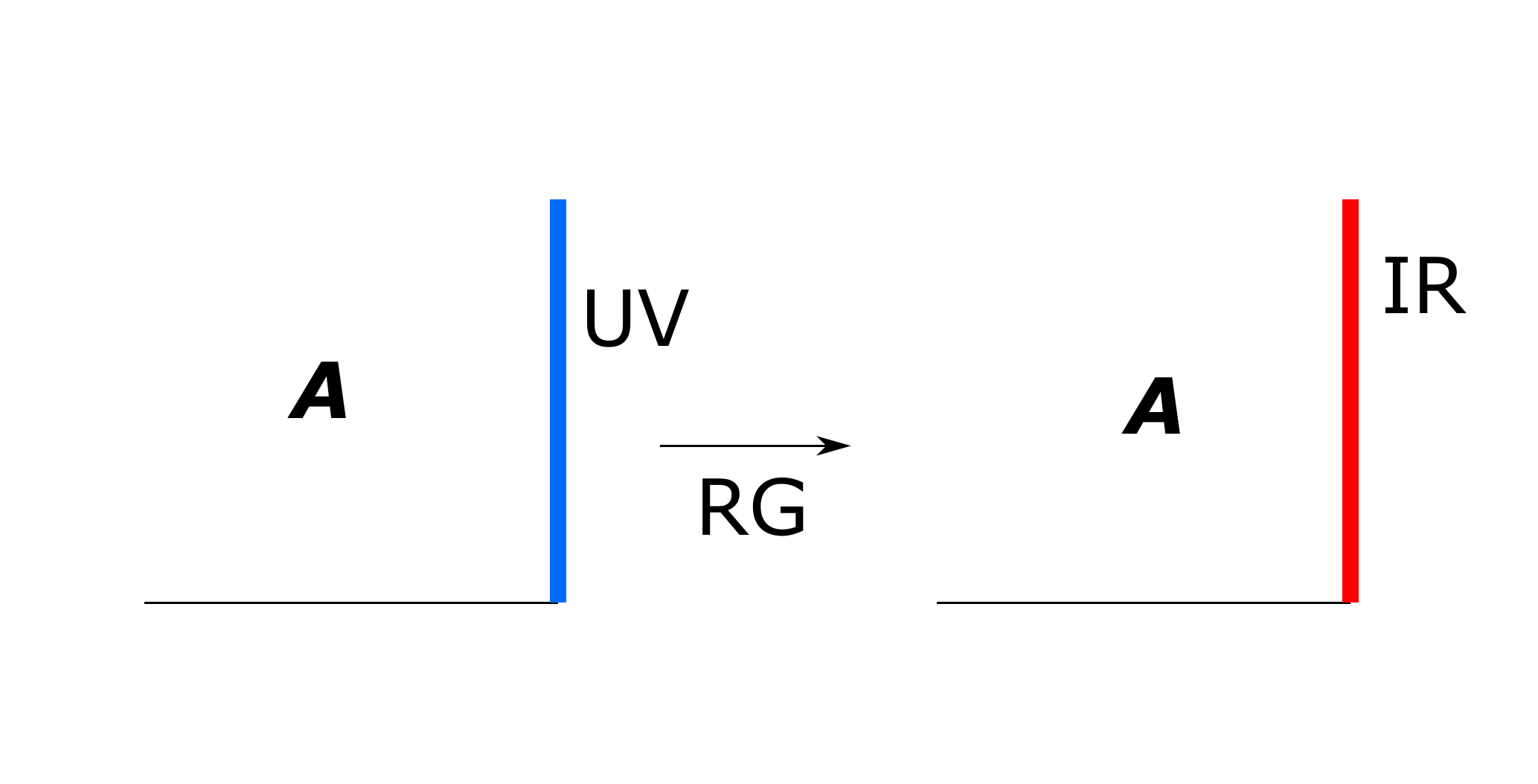}&
    \includegraphics[scale=0.35]{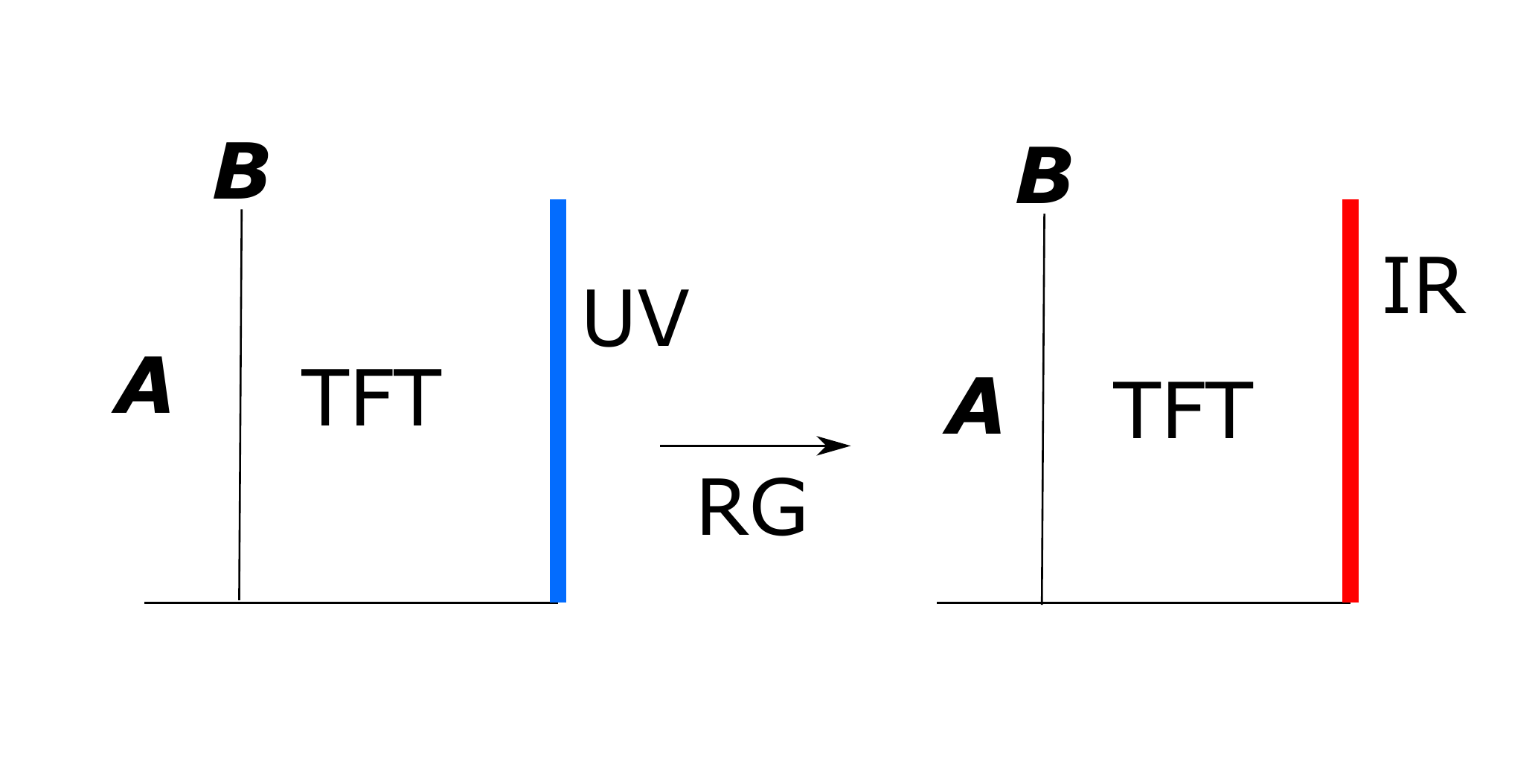}
    \end{tabular}
    \caption{\textsc{Left}: Inflow picture of anomaly matching for intrinsic QFTs: the anomaly is an invertible TFT $\boldsymbol{A}$ in one dimension higher. \textsc{Right}: TFT matching for QFTs with global structure. The choices of global structure are encoded by non-invertible TFT on an interval with a boundary condition interface $\boldsymbol{B}$ to an anomaly theory. This symmetry TFT must match for the theories in the UV and the IR, thus generalizing the anomaly matching procedure.}
    \label{fig:TFTmatching}
\end{figure}

In the presence of anomalies for continuous symmetries the symmetry
theory includes sectors of Chern-Simons type, or more generally
$\eta$-invariants. In this note we focus on the choice of global form,
which involves discrete symmetries only, so we do not need to worry
about such sectors, and we can restrict ourselves to a part of the
symmetry theory which is a proper topological field theory (TFT). By
an abuse of language, we will refer to this TFT sector of the symmetry
theory as \emph{symmetry TFT}. It is natural to expect this symmetry
TFT to be invariant under RG flows triggered by deformations invariant
under the symmetries. Equivalently, we expect that the set of choices
for the global form of the theory persists at all energy scales ---
see figure~\ref{fig:TFTmatching}.

\medskip

We will test this expectation on four dimensional QFTs in four space
time dimensions that admit a Coulomb phase, i.e. an infrared regime
with an effective description in terms of $r$ independent Maxwell
fields. Let us denote such theory $\mathcal T$. In this paper we are
interested in determining the global structure of $\mathcal T$ from
the perspective of such infrared regime. All the examples we will
consider in this note are $\cN=2$ supersymmetric, but we stress that
supersymmetry in itself is not a necessary requirement for the
discussion below: the two assumptions we are making on $\mathcal T$
are that it has an infrared regime where an effective $U(1)^r$ gauge
theory description is valid, and that the structure of the massive
spectrum in this IR regime is sufficiently well understood. The
restriction to $\cN=2$ comes from this second assumption, which in the
$\cN=2$ context we take to mean the (likely weaker) assumption that
the BPS spectrum, which we understand well enough for our needs, is
representative of the full spectrum.\footnote{We emphasize that this
  as an assumption. A known example where the non-BPS spectrum
  includes states that would invalidate an analysis based on the BPS
  spectrum only is the type I string, with gauge group
  $\Spin(32)/\bZ_2$. In this case the spinor representation appears as
  a non-BPS brane with discrete K-theory charge; the BPS spectrum
  appears in vector representations only. In all examples without
  torsional charges we know the BPS spectrum is representative. It is
  tempting to conjecture that this is a general fact.}

\medskip

The class of $\cN=2$ theories that we study naturally includes
geometrically engineered $\cN=2$ SCFTs. In the geometric engineering
program \cite{Katz:1996fh} one aims to establish a dictionary between the properties of
some version of string theory, denoted by $\scS$, on a singular
background\footnote{\ Here we let $\mathscr S$ denote also M-theory or
  F-theory, and by $\mathcal X$ we denote schematically the whole data
  needed to prescribe a BPS background for $\mathscr S$ of the form
  $\mathbb R^{d} \times \mathcal X$ giving rise to a QFT in $d$
  spacetime dimensions.} $\cX$ and a quantum field theory
$\mathcal T_{\mathscr S/\mathcal X}$. From this perspective the
symmetry TFT can be recovered via the analysis of the effective theory
arising after compactification on $\partial\cX$, which captures the
anomalies and other features of $\mathcal T_{\mathscr S/\mathcal
  X}$, including its higher form symmetries in terms of defect groups \cite{DelZotto:2015isa,Heckman:2017uxe,Eckhard:2019jgg,GarciaEtxebarria:2019caf,Albertini:2020mdx,Morrison:2020ool,Dierigl:2020myk,Closset:2020scj,DelZotto:2020esg,Apruzzi:2020zot,Bhardwaj:2020phs,Closset:2020afy,Heidenreich:2020pkc,DelZotto:2020sop,Gukov:2020btk,Bah:2020uev,Bhardwaj:2021pfz,Apruzzi:2021mlh,Apruzzi:2021phx,Hosseini:2021ged,Apruzzi:2021vcu,Bhardwaj:2021wif,Bhardwaj:2021zrt,Closset:2021lwy,Heidenreich:2021xpr,Buican:2021xhs,Cvetic:2021maf, Debray:2021vob,Apruzzi:2021nmk,Braun:2021sex,Bah:2021brs,Bhardwaj:2021mzl,Cvetic:2020kuw,Cvetic:2021sxm,Apruzzi:2022dlm,Hubner:2022kxr,DelZotto:2022joo,Cvetic:2022imb}. In particular, if the theory
$\mathcal T_{\mathscr S/\mathcal X}$ has non-trivial choices of global
structure, in the simplest cases this has been understood \cite{Garcia-Etxebarria:2019cnb,Albertini:2020mdx,DelZotto:2020esg,Bhardwaj:2021mzl} in terms of
a Heisenberg algebra of non-commuting FMS fluxes on $\partial \mathcal
X$ \cite{Freed:2006ya,Freed:2006yc}. Below we will give an alternative purely field
theoretical derivation of this same Heisenberg algebra, showing that
it also arises from the infrared perspective on theories with a
Coulomb phase. The advantage of this formulation is that it extends
field theoretically the results about theories with known Lagrangian
formulations to arbitrary SCFTs with mutually non-local massless
excitations. In particular, we recover field theoretically the results
on global structures of four-dimensional Argyres-Douglas theories that
have been obtained via geometric string theory techniques in the
literature, and also some results that have no geometric
understanding.

\medskip

Our main result can be derived closely following
\cite{Aharony:2013hda}.
In the pure $\mathfrak{u}(1)^r$ gauge theory, before choosing a global
form, one can in principle consider Wilson and 't Hooft lines with
arbitrary rational dyonic charges. Once we include massive states, two
things happen: the set of allowable charges for the lines reduces to
those mutually local with respect to the charges of the dynamical
states, and some of the line operators get screened. Depending on the
structure of the charge lattice of the theory non-trivial lines might
remain after screening. Generically not all remaining lines will be
mutually local, so an specification of a global form will consist, as
in \cite{Aharony:2013hda}, on a specification of a maximal subset of
lines that will be genuine line operators in our theory. The rest of
the lines should then be viewed as open surface operators.

\medskip

We emphasise that from this point of view the usual choice of having
line operators with arbitrary integer dyonic charge is a possible
choice for the maximally commuting set of line operators, but as shown
below this is not the only possible choice given any fixed lattice of
charged states. (In making this statement we assume that we have
normalised our charges so that the lattice of charged states is a
sublattice of $\bZ^{2r}$, which will be the case throughout the
paper.)

\medskip

The argument in terms of screening of lines given above can be recast
as the derivation of a symmetry TFT for the theory of the
form\footnote{Throughout the paper we will use conventions where
  the $b$ fields are periodic with period $1$, instead of $2\pi$.}
\begin{equation}
S_{5d} = 2\pi i\sum_{j=1}^r {  n_j} \int_{\mathcal M_5} b^{(2)}_{e,j} \wedge d b^{(2)}_{m,j}\,
\end{equation}
where $n_j \in \mathbb Z_{\geq 1}$. Below we will give an explicit
prescription for how to extract the integers $n_j$ from the Coulomb
phase of the theory. When some of these integers differ from 1 we
recover the Heisenberg algebra of non-commuting fluxes we found from
geometric engineering in terms of this bulk TFT.

\medskip

All conventional 4d $\cN=2$ gauge theories can be analysed with our
methods: giving generic vevs in the Cartan of the gauge group to the
adjoint scalars in the $\cN=2$ vector multiplets, breaks the gauge group
to $U(1)^r$, giving us a plethora of consistency checks. Moreover,
this feature is also shared by all non-conventional 4d $\cN=2$ SQFTs
that have a Coulomb branch of complex dimension $r$. In this case the
integer $r$ is known as the `rank' of the corresponding
non-conventional SQFT. From this perspective our results extend and
generalize results that have appeared previously in the literature by
giving a common ground for many computations of defect groups in
various dimensions.

\medskip

We conclude this paper by extending our formulation to other theories with IR phases under control, namely 
6d SCFTs with a tensor branch, and 5d SCFT with a Coulomb branch. Also for those systems we find a symmetry TFT that has the structure of a BF theory in 7 ad 6 dimensions respectively, which is responsible for the global structure of the theories in question.

\medskip

The structure of this paper is as follows. In section
\ref{sec:heureka} we review some well-known features of Maxwell type
theories and their higher form symmetries to set up our notation and
conventions. We proceed by revisiting the field theoretical origin of
global structures for 4d theories from an infrared perspective. In
section \ref{sec:TFT} we summarise the main features of BF theories we
will need, and derive our main result. Results in these two
sections only assume the theory under scrutiny has a Coulomb phase,
and are independent from supersymmetry. In section \ref{sec:examples}
however, we apply this result in the context of $\mathcal N=2$
theories where the charge lattice of the theories can be explicitly
calculated, from this we recover the well-known center symmetries and
global structures of conventional gauge theories, and we also
reproduce and extend the results obtained about non-conventional SCFTs
via geometric engineering methods. In section \ref{sec:generaldims} we
give a generalization of our findings to theories in various
dimensions. In section \ref{sec:conclusions} we present our
conclusions and comment on future directions and applications.

\section{Genuine and non-genuine lines from the infrared}\label{sec:heureka}

\subsection{Line operators from the Coulomb phase}

In what follows we explore the constraints on the line operators of
Maxwell theory that arise from the presence of charge dyonic BPS
states in the spectrum. Many important aspects of the analysis below
can be found in section 4.1 of \cite{Gaiotto:2014kfa}, see also
appendix C in \cite{Cordova:2018cvg}.

\medskip

\begin{figure}
    \centering
    \includegraphics[scale=0.5]{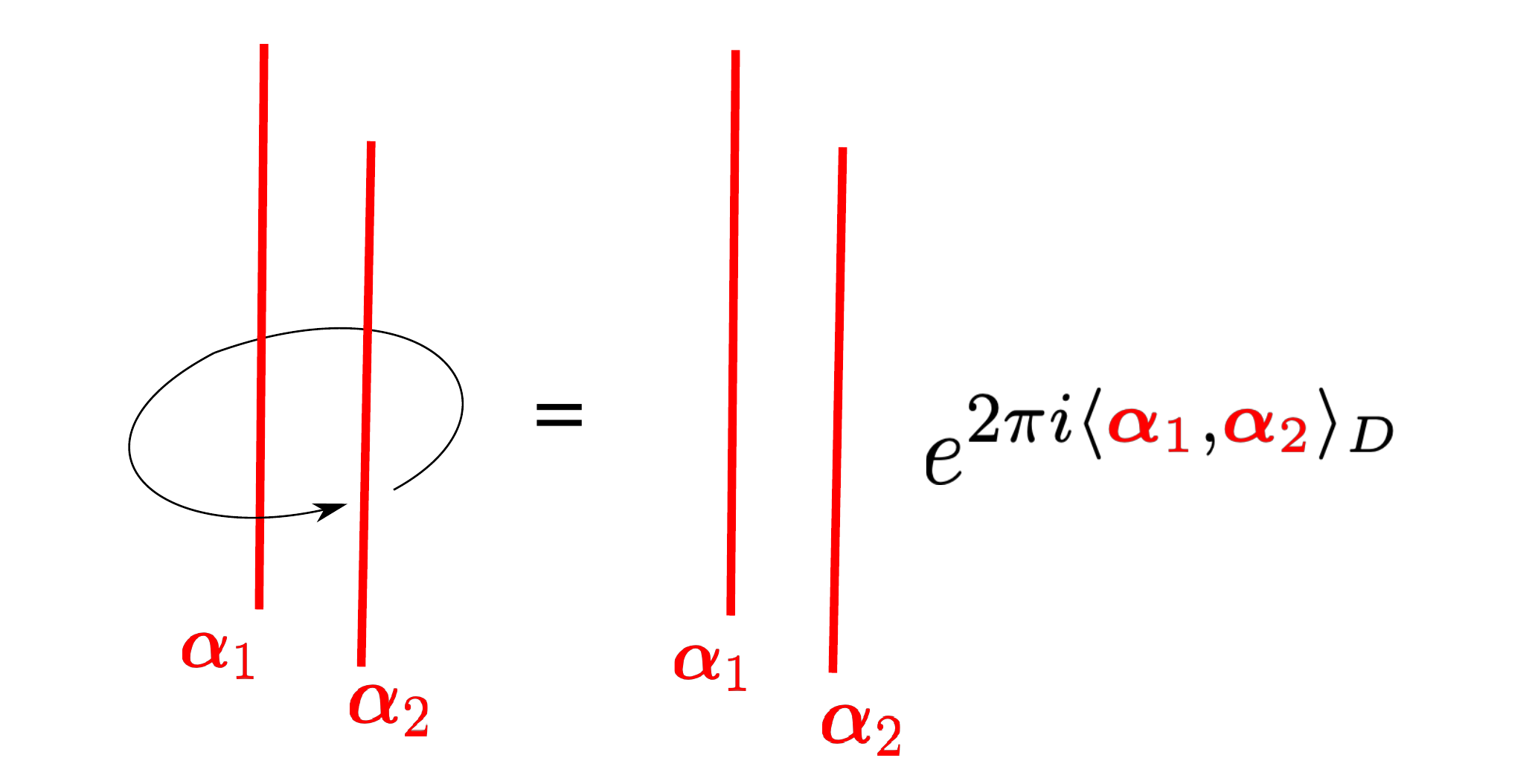}
    \caption{Aharonv-Bohm effect for lines: monodromies can give rise
      to phases proportional to the corresponding Dirac pairings. Only
      lines which have charges satisfying the Dirac quantization
      condition can be simultaneously genuine. The presence of
      non-genuine lines is the hallmark of a relative QFT -- see
      e.g. section 2 of \cite{Bhardwaj:2021mzl} for a nice review.}
    \label{fig:nongenuinelines}
\end{figure}

The dynamics of a theory in a Coulomb phase can be described in terms
of $r$ copies of the $\fu(1)$ Maxwell theory coupled to massive
states. (There can be non-generic points where some of these massive
states can become massless.) Ignoring the massive states for a moment,
at a generic point we have a higher 1-form global symmetry of the form
\begin{equation}\label{eq:naive}
\big(U(1)^{(1)}_e \times U(1)^{(1)}_m\big)^r
\end{equation}
where $U(1)^{(1)}_e \times U(1)^{(1)}_m$ are the 1-form global
symmetries of Maxwell theory. The operators charged under these
symmetries are dyonic lines. We denote the electric and magnetic
charges of one such line by a $2r$-component vector
$\boldsymbol{\alpha}$. As in \cite{Gaiotto:2010be,Aharony:2013hda},
two lines with charges $\boldsymbol{\alpha}_1$ and
$\boldsymbol{\alpha}_2$ can only be genuine operators in the Maxwell
theory (as opposed to open Gukov-Witten surface operators
\cite{Gukov:2006jk,Gukov:2008sn}) if their Dirac pairing
$\dsz{\boldsymbol{\alpha}_1}{\boldsymbol{\alpha}_2}_D\df
\sum_{i=1}^r(\boldsymbol{\alpha}_1^{2i-1}\boldsymbol{\alpha}_2^{2i} -
\boldsymbol{\alpha}_1^{2i}\boldsymbol{\alpha}_2^{2i-1})$ is an
integer.\footnote{Relatedly, we can determine the commutation
  relations of the line operators in a spatial slice using the
  exponentiated form of the commutators found in
  \cite{Ashtekar:1997rg}, which leads to a commutator
  $\exp(2\pi i
  \dsz{\boldsymbol{\alpha}_1}{\boldsymbol{\alpha}_2}_D)$.} (See
figure~\ref{fig:nongenuinelines}.) This is often accomplished by
requiring $\boldsymbol{\alpha}_1,\boldsymbol{\alpha}_2\in\bZ^{2r}$,
but in the absence of a dynamical spectrum of point particles this is
not necessary for consistency of the theory.

\medskip

\begin{figure}
\begin{center}
\includegraphics[scale=0.6]{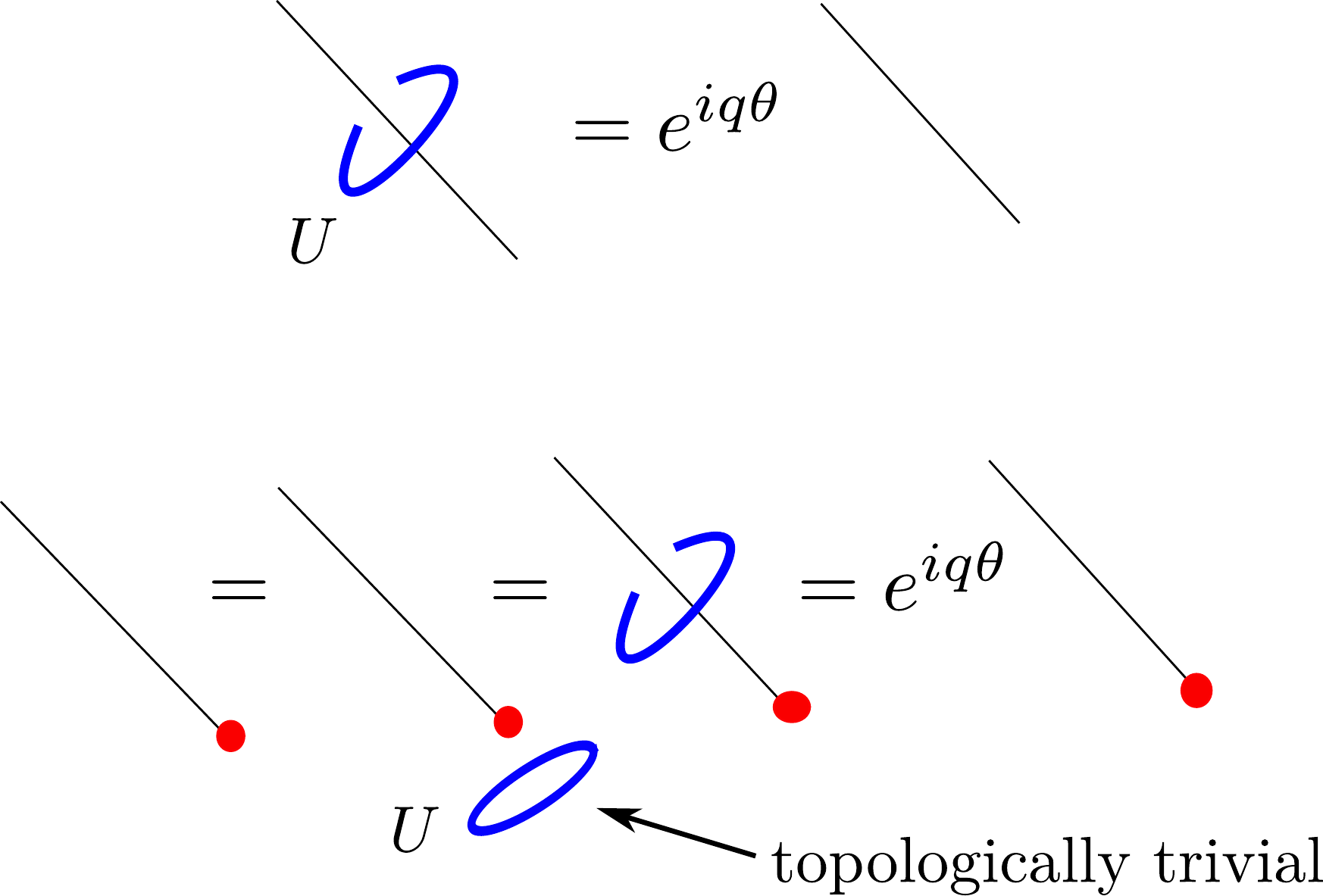}
\end{center}
\caption{\textsc{Top:} action of a $U(1)^{(1)}$-form symmetry on a
  line operator. \textsc{Bottom:} Constraint on the higher form
  symmetry in presence of charged particles (red dot at the end of the
  line). The presence of a particle of charge $q\in\mathbb Z$ enforces
  $e^{i\theta q} = 1$ which in turns entail that only rotations with
  phase $\theta = 2 \pi k / q$ are allowed, thus breaking $U(1)^{(1)}$
  down to $\mathbb Z^{(1)}_q$.}\label{fig:boilingwater}
\end{figure}

This structure gets simplified once we couple the pure Maxwell theory
to charged dynamical states. We will refer to this theory as the ``UV
theory'', in contrast to the ``IR theory'' by which we mean pure
Maxwell with no dynamical states. The electromagnetic charges of the
dynamical states live on a lattice $\Gamma\subseteq \bZ^{2r}$. In
practice, it is convenient to choose a basis $\gamma_i$ of generators
for $\Gamma$, and this gives rise to an explicit expression for the
Dirac pairing in terms of
$\mathcal Q_{ij} = \langle \gamma_i, \gamma_j\rangle_D\in\bZ$.

The first effect of coupling the theory to charged dynamical states is
that the spectrum of admissible lines is reduced. Given a line with
charges $\boldsymbol{\alpha}$, it is only admissible if
\begin{equation}
  \label{eq:Diracchio}
  \dsz{\boldsymbol{\alpha}}{\gamma}_D\in\bZ \quad \text{for all } \gamma\in\Gamma\,.
\end{equation}
In other words, $\boldsymbol{\alpha}\in\Gamma^*$, the dual lattice.
The lines in $\Gamma^*$ do not need to commute --- see figure
\ref{fig:nongenuinelines}. Integrally charged line operators always
commute, as pointed out in \cite{Freed:2006yc}, but $\Gamma^*$ is not
necessarily an integral lattice.  This is the Coulomb branch
counterpart of the effect discussed at length by
\cite{Gaiotto:2010be,Aharony:2013hda}: the same local dynamics can be
compatible with inequivalent global structures, and different global
forms of the theory can be detected from the spectrum of line defects
in four-dimensions.

Moreover, the spectrum of allowed charges for genuine line defects is
also constrained as a sublattice of $\Gamma_L \subset \Gamma^*$
consisting of a maximally mutually local collection of defect charges
satisfying the Dirac quantization condition
$\langle \boldsymbol{\alpha}, \widetilde{\boldsymbol{\alpha}}\rangle_D
\in \mathbb Z$. Inequivalent global forms of the theory correspond to
inequivalent choices of sublattices $\Gamma_L$ after screening.

\medskip

Indeed, a second effect is that the higher form symmetry
group~\eqref{eq:naive} is explicitly broken to a subgroup via
screening. 
This can be understood as follows: the line defects can have endpoints
corresponding to charged operators, and this constrains the electric
and magnetic symmetry operators in the UV theory to be those under
which lines which can end are neutral. For an example see
figure~\ref{fig:boilingwater} where we illustrate the breaking of a
$U(1)^{(1)}$ higher symmetry to ${\mathbb Z_q}^{(1)}$ in presence of a
particle of charge $q\in\mathbb Z$. From this perspective, the
symmetry \eqref{eq:naive} is an emergent symmetry in the IR, which is
broken by massive states that at sufficiently high energies become
dynamical. For this reason the UV theory has a much smaller 1-form
global symmetry, which oftentimes is completely trivial (corresponding
to the cases when the spectrum of the theory is complete, meaning that
all defect charges can be screened).\footnote{ For a recent discussion
  about this point and applications beyond the scope of the present
  work, we refer our readers to the nice work
  \cite{Rudelius:2020orz}.} A related effect is that the potential
background fluxes that the theory admits get reduced: a background for
$U(1)_e$ can be understood as a modification for the quantisation
condition for the magnetic flux (see e.g. section 5 of
\cite{Hsin:2019fhf} for a discussion), so in the presence of
electrically charged states only those backgrounds that result in
quantisation conditions compatible with the charges of the dynamical
matter are allowed. For instance, if all of our electrically charged
dynamical particles have even charge, we can only have background
magnetic fluxes with holonomies 0 and $\frac{1}{2}$. In general, the
holonomies of the electric and magnetic fluxes must be such that their
Dirac pairing with all dynamical states is integral.

The screening of $\Gamma^*$ with respect to $\Gamma$ gives rise to
1-form factor of the defect group, $\mathbb D^{(1)} = \Gamma^*/\Gamma$
\cite{DelZotto:2015isa}. The actual higher form symmetry for a given
theory corresponding to the sublattice $\Gamma_L$ is
$\Gamma_L/\Gamma$ \cite{Gaiotto:2010be,Aharony:2013hda}.

\subsection{An example: the $\cN=2$ $\mathfrak{su}(2)$ theory on the Coulomb branch}
\label{sec:su(2)}

As a simple illustration of the previous discussion, here we briefly
review the results of \cite{Gaiotto:2010be} about the global form of
the $\mathfrak{su}(2)$ gauge theories. From the Seiberg-Witten
solution it is known that the BPS spectrum of the $\mathfrak{su}(2)$
theory can be generated as bound states of two mutually non-local
excitations with charges $\gamma_1$ and $\gamma_2$ such that
$\langle \gamma_1, \gamma_2 \rangle_D = - 2$. In the choice of
electromagnetic duality frame by Seiberg and Witten one can view
$\gamma_1$ as a monopole of charge $(0,1)$ and $\gamma_2$ as a dyon of
charge $(2,-1)$. To determine the dual lattice $\Gamma^*$ we can
proceed as follows
\begin{equation}
    \langle\boldsymbol{\alpha},\gamma_1\rangle_D = \alpha_1 \qquad \langle\boldsymbol{\alpha},\gamma_2\rangle_D = -\alpha_1 - 2\alpha_2
\end{equation}
Then the first equation implies that $\alpha_1 \in \mathbb Z$, while
the second implies that $\alpha_2 \in {1\over 2}\mathbb Z$. Let us
consider the possible mutually local sublattices of $\Gamma^*$ from
this perspective, with respect to the corresponding screenings. There
are three minimal defect charges which get nontrivial monodromies with
respect to one another, corresponding to non-integer quantized Dirac
pairings, namely
\begin{equation}
(1,0) \qquad (0,\tfrac{1}{2}) \qquad\text{and}\qquad (1,\tfrac{1}{2})\,.
\end{equation}
Let us first consider the sublattice $\Gamma^*_{(1,0)} \subset \Gamma^*$ that contains the line defect with charge $(1,0)$. The requirement of maximal mutual locality then implies that
\be
\langle (1,0), \boldsymbol{\alpha}\rangle_D = \alpha_2 \in \mathbb Z
\ee
Hence the charges of the line defects in $\Gamma^*_{(1,0)}$ have the form $(n,m)$ where $n,m$ are both integers. Considering the screening by the charges $\gamma_1$ and $\gamma_2$ corresponds to identifying
\begin{equation}
    (n,m) \sim (n',m') + k_1 \gamma_1 + k_2 \gamma_2
\end{equation}
where $k_i \in \mathbb Z$. We see that we are left only with two
equivalence classes in $\Gamma^*_{(1,0)}$, namely $[(1,0)]$ and
$[(0,0)]$, hence we obtain an electric 1-form symmetry
$(\mathbb Z_2)^{(1)}_e$ corresponding (with a natural choice of
duality frame) to the gauge group $SU(2)$. Now consider the lattice
$\Gamma^*_{(0,\tfrac{1}{2})}$ which contains the line with charge
$(0,\tfrac{1}{2})$. The requirement of maximal mutual locality for
this class of charges is
\begin{equation}
    \langle (0,\tfrac{1}{2}), \boldsymbol{\alpha}\rangle_D = - {\alpha_1\over 2} \in \mathbb Z \, \Rightarrow \, \alpha_1 \in 2\mathbb Z
\end{equation}
Hence the charges of the line defects in $\Gamma^*_{(0,\tfrac{1}{2})}$ have the form $(2n,m/2)$ where $n,m$ are both integers. The screening equivalence relation is again
\begin{equation}
    (2n,m/2) \sim (2n',m'/2) + k_1 \gamma_1 + k_2 \gamma_2.
\end{equation}
We see that we are left with only two equivalence classes again
$[(0,0)]$ and $[(0,{1\over 2})]$ corresponding to a magnetic 1-form
symmetry $(\mathbb Z_2)^{(1)}_m$ which gives a gauge group
$SO(3)_+$. Now, consider
$\Gamma^*_{(1,\tfrac{1}{2})}$: procceeding analogously, we obtain that
\begin{equation}\begin{aligned}
\langle (1,\tfrac{1}{2}), \boldsymbol{\alpha}\rangle_D &= \alpha_2 - {\alpha_1\over 2} \in \mathbb Z \\ &\Rightarrow \, (\alpha_1,\alpha_2) \in \{((2n+1),(2m+1)/2)\} \text{ or }  \{(2n,m)\} \qquad n,m\in\mathbb Z 
\end{aligned}
\end{equation}
Again by screening we see that we obtain only two equivalence classes: $[(0,0)]$ and $[(1,\tfrac{1}{2})]$, corresponding to a $(\mathbb Z_2)^{(1)}_{diag}$ 1-form symmetry, which corresponds to the gauge group $SO(3)_-$.

\medskip

We stress here that the above result is independent of the choice of electromagnetic frame: we can choose to work with any different basis as long as we are preserving the Dirac paring. For instance, one could do the analysis working with $\gamma_1 = (1,1)$ and $\gamma_2 = (1,-1)$. In this case the charges in $\Gamma^*$ have the form
\begin{equation}
    \langle\boldsymbol{\alpha},\gamma_1\rangle_D = \alpha_1 - \alpha_2 \qquad \langle\boldsymbol{\alpha},\gamma_2\rangle_D = - \alpha_1 -\alpha_2
\end{equation}
And therefore one obtains
\begin{equation}
    (\alpha_1,\alpha_2) \in \{(\tfrac{2m+1}{2},\tfrac{2n+1}{2})\} \text{ or } \{(m,n)\} \qquad m,n \in \mathbb Z.
\end{equation}
With this choice of duality frame we have the following defect charges that would violate Dirac quantization
\begin{equation}
    (\tfrac{1}{2},\tfrac{1}{2}) \qquad (\tfrac{1}{2},-\tfrac{1}{2}) \qquad\text{and}\qquad (0,1)\,.
\end{equation}
These correspond to the three choices of lattices above, if we identify the direction $(1,1)$ with the magnetic charge and the direction $(0,1)$ with the electric one.

\section{Symmetry TFT and global structure}\label{sec:TFT}

We have just argued that whenever the lattice $\Gamma$ of charges
particles in Maxwell theory is not unimodular we have the possibility
of having choices of global structure, encoded as choices of maximal
sets $\Gamma_L$ of commuting elements in $\Gamma^*/\Gamma$. This is
perhaps a surprising statement, as we typically don't think of Maxwell
theory as admitting different global forms. The key difference between
our analysis and the usual analysis is that we do not impose a priori
integral quantization for the electric and magnetic fluxes in the
theory, but rather accept as valid any flux quantisation structure
compatible with the dynamical matter content. For instance, if all
electrically charged states have even charge we include
half-integrally quantised fluxes in the path integral for Maxwell
theory. If we are considering theories with only electrically charged
states (in some duality frame) then this is purely a matter of
convention, and the half-integrality can be rescaled away.  In
contrast, the theories of interest to us are richer, and include
dyonic states, which lead to genuinely different prescriptions for
which fluxes to sum over.

Our task is therefore classifying all the possibilities for flux
quantisation conditions compatible with the local dynamics. As
mentioned above, the choice of quantisation for the electric and
magnetic fields can be understood as a choice of background fields for
the electric and magnetic 1-form symmetries. So our problem may be
recast as the determination of which choices of background fluxes are
allowed in a given quantum theory, given the dynamical matter
content. This kind of problem has a familiar solution (see for
instance \cite{Witten:1998wy}): the possible flux choices can be
understood as states in the Hilbert space of a (generically
non-invertible) BF theory in one dimension higher.

\subsection{A quick review of BF theory in $D+1$ dimensions}\label{app:BFreview}
In this section we review some basic details of the BF theory,
following the discussion in \cite{Kapustin:2014gua}. These details are
well-known, and can be skipped by \textit{cognoscenti}. A BF theory is
a TFT in $D+1$ dimensions with action
\begin{equation}
\begin{aligned}
  S = {2\pi i n} \int_{\mathcal M^{D+1}} b^{(q+1)} \wedge d b^{(D-q-1)}
\end{aligned}
\end{equation}
where $b^{(q+1)}$ and $b^{(D-q-1)}$ are a $(q+1)$-form and a $(D-q-1)$-form. We stress that the coefficient $n$ which multiplies the action must be an integer for $\text{exp}(- S)$ to be well defined and compatible with the local $U(1)$ gauge transformations 
\begin{equation}
  b^{(q+1)} \to b^{(q+1)} + d \lambda^{(q)} \qquad b^{(D-q-1)} \to b^{(D-q-1)} + d \lambda^{(D-q-2)}\,.
\end{equation}
Here we are being naive and focusing only on the local structure of
these higher gauge transformations. More refined experts might look
into the full fledged gerby behavior of these Deligne-Beilinson
cocylces. For our purposes in this paper the above description will
suffice. Notice that upon a gauge variations we obtain
\begin{equation}
  S \to S + {2\pi i n}  \int_{\partial \mathcal M^{D+1}} \lambda^{(q)} \wedge d b^{(D-q-1)}\,.
\end{equation}
This boundary term is crucial for applications to inflow and generalizations that we are after in this paper.

\medskip

The theory is topological for a simple reason: the equation of motion
for this theory are
\begin{equation}\label{eq:BFeom}
f^{(q+2)} = d b^{(q+1)} = 0 \qquad\qquad f^{(D-q)} = d b^{(D-q-1)} = 0
\end{equation}
and forbid any local propagating degree of freedom. The theory has
nevertheless interesting non-local gauge invariant operators
corresponding to closed $(q+1)$-dimensional and $(D-q-1)$-dimensional
hypersurfaces of $\mathcal M^{D+1}$, that generalize the familiar
Wilson lines:
\begin{equation}
\begin{aligned}
&\mathcal W_{\Sigma^{(q+1)}} = \exp 2\pi i \int_{\Sigma^{(q+1)}} b^{(q+1)} \\ 
&\mathcal W_{\Sigma^{(D-q-1)}} = \exp 2\pi i \int_{\Sigma^{(D-q-1)}} b^{(D-q-1)} 
\end{aligned}
\end{equation}
To determine the algebra of these operators notice that the insertion
of $\mathcal W_{\Sigma^{(q+1)}}$ in a correlator can be absorbed in
the action introducing a source term in the equations of motion
\eqref{eq:BFeom}
\begin{equation}
  n f^{(D-q)} = \delta_{\Sigma^{q+1}}
\end{equation}
where $\delta_{\Sigma^{(q+1)}}$ is the Poincar\'e dual to the cycle $\Sigma^{(q+1)}$ in $\mathcal M^{D+1}$. As a result we obtain that
\begin{equation}
\langle \mathcal W_{\Sigma^{(q+1)}}  \mathcal W_{\Sigma^{(D-q-1)}} \rangle = \exp \left( {2\pi i \over n}  \ell(\Sigma^{(q+1)},\Sigma^{(D-q-1)}) \right)
\end{equation}
where $ \ell(\Sigma^{(q+1)},\Sigma^{(D-q-1)}) $ is the linking number of $\Sigma^{(q+1)}$ and $\Sigma^{(D-q-1)}$ in $\mathcal M^{D+1}$. Equivalently, restricting everything along a spatial slice and considering a Hamiltonian quantization, these generalized Wilson lines form a Heisenberg algebra
\begin{equation}
\mathcal W_{\Sigma^{(q+1)}}  \mathcal W_{\Sigma^{(D-q-1)}} = \exp  {2\pi i \over n} \left( \Sigma^{(q+1)} \cdot \Sigma^{(D-q-1)} \right)  \mathcal W_{\Sigma^{(D-q-1)}}  \mathcal W_{\Sigma^{(q+1)}}  
\end{equation}
where $\cdot$ is the intersection pairing along the spatial slice for
the Hamiltonian quantization. Since this Heisenberg algebra is
nontrivial
the Hilbert space associated to a generic codimension one submanifold
$\cM^D$ will have dimension greater than one.

\subsection{Global structures from the infrared}

We now specialize the discussion in the previous section to the case
of interest for this paper in which $D = 4$ and $q=1$.
Our main claim is that the introduction of the dynamical states on
Maxwell theory leads to a 4d theory relative to the 5d
theory\footnote{We stress that this is not the whole action of the
  symmetry TFT, just the part from which the choice of global
  structures will follow.}
\begin{equation}
  \label{eq:BF-Q}
  \cS = {2\pi i \over 2} \int_{\mathcal M^5}  \mathcal Q^{\alpha \beta} \,  b^{(2)}_{\alpha} \wedge d b^{(2)}_{\beta} = \frac{2\pi i}{2} \int_{\cM^5} \dsz{b^{(2)}}{db^{(2)}}_D
\end{equation}
where the matrix $\mathcal Q^{\alpha\beta}$ is the $2r\times 2r$ Dirac
pairing for the BPS states in 4d, and $b^{(2)}$ is a $2r$ component
2-form (or more precisely, a $2r$-dimensional vector of degree 3
differential characters, but as mentioned above we will not need to
worry about such topological subtleties here).

This action can be justified as follows. Note first that since $\cQ$
is an antisymmetric matrix we can do an invertible integral change of
basis to bring it into a block diagonal form
\begin{equation}
  \label{eq:tilde-Q}
  \widetilde{\cQ} = \left(\begin{array}{cccccccc|ccc} 0 & n_1 &  &  &  \cdots  &  &  &  \\  - n_1 & 0 &  &  &  & & &  \\  &&\ddots &&& && \vdots \\ \vdots & & & 0  & n_k && & \\ & & & - n_k & 0 &&& & \\  & & &  &  & \ddots &&   \\  & & &  &  &  &  0 & n_r \\  &  & & \cdots  &  &  & -n_r &  0 \\\hline && &&& && &  \\ && &&& && & & \mathbf{0}_{f \times f} \\   && &&& && & & & \end{array}\right)
\end{equation}
with the $n_i$ integral and non-negative (see theorem IV.1 of
\cite{newman1972integral} as well as the discussion in \cite{Caorsi:2017bnp}). In practice the $n_i$ can be obtained
easily by going to the Smith normal form of $\cQ$. If we name the
components of
$b^{(2)}=(b^{(2)}_{e,1}, b^{(2)}_{m,1}, b^{2}_{e,2},\ldots)$ we see
that our 5d theory becomes (up to boundary counterterms that we are
neglecting)
\begin{equation}
  \label{eq:sector}
  \mathcal S_5 = {2\pi i} \sum_{i=1}^r n_i  \int_{\mathcal M^5}  b^{(2)}_{e,i} \wedge d b^{(2)}_{m,i}\, .
\end{equation}
So our 5d theory decomposes into a sum of decoupled sectors.

In order to justify~\eqref{eq:sector}, consider the effect of going to
the basis giving~\eqref{eq:tilde-Q} on the boundary field theory. The
generators $\gamma_i$ in this basis satisfy
$\dsz{\gamma_{2i-1}}{\gamma_{2i}}_D =
-\dsz{\gamma_{2i}}{\gamma_{2i-1}}_D = n_i$ and zero otherwise. So the
problem reduces to a situation similar to what we found in the
$\fsu(2)$ case in \S\ref{sec:su(2)}. For simplicity we will henceforth
focus on the first block.

By an $SL(2,\bZ)$ transformation we can further choose
$\gamma_1=(0,1)$, $\gamma_2=(n_1,k)$, with $k\in\bZ$. We see that all
states in this basis have electric charge divisible by $n_1$, so we
can consistently introduce 't Hooft lines of charge $1/n_1$. Wilson
lines, on the other hand, necessarily have integral charge. More
generally, the spectrum of allowed lines has charges $(p,q/n_1)$, with
$p,q\in\bZ$. This implies, in turn, that the electric background
fields have periodicity 1, but magnetic background fields have
periodicity $n_1$. This means that the allowed flux operators on the
5d BF theory are of the form $\cW_e^{p}\cW_m^{q/n_1}$, with
$p,q\in\bZ$ (but $n_1$ not necessarily dividing $q$). Equivalently, we
can think of the line operators as being the boundary of Gukov-Witten
open surface operators, which when pulled to the 5d bulk become the
operators in the BF theory.

A BF theory with action
\begin{equation}
  \cS_{5} = 2\pi i \int_{\cM^5} b_e^{(2)}\wedge db_m^{(2)}
\end{equation}
where $b_m^{(2)}$ has periodicity $n_1$ (so $\cW_m^{1/n_1}$ is
allowed) and $b_e^{(2)}$ periodicity 1 is the same as a BF theory with
action
\begin{equation}
  \cS_{5} = 2\pi i\, n_1 \int_{\cM^5} b_e^{(2)}\wedge d\tilde b_m^{(2)}
\end{equation}
with both $b_e^{(2)}$ and $b_m^{(2)}$ of periodicity 1 (via
$b_m^{(2)}\df n_1\tilde b_{m}^{(2)}$). Unwinding the choices of basis,
this proves that~\eqref{eq:BF-Q} is indeed the bulk theory for Maxwell
theory with pairing $\cQ$.





From our review in section \ref{app:BFreview} it follows that this TFT
has an algebra of generalized Wilson lines of the form
\begin{equation}
\mathcal W^{e,i}_{\Sigma^2} = \exp 2\pi i \int_{\Sigma_2} b^{(2)}_{e,i}  \qquad \mathcal W^{m,i}_{\Sigma^2} = \exp 2\pi i \int_{\Sigma_2} b^{(2)}_{m,i}
\end{equation}
which along a spatial slice satisfy an Heisenberg algebra
\begin{equation}
\mathcal W^{e,i}_{\Sigma^2} \mathcal W^{m,i}_{\widehat{\Sigma}^2} = \exp\left( {2\pi i \over n_i} (\Sigma^2 \cdot \widehat{\Sigma}^2)\right) \mathcal W^{m,i}_{\widehat{\Sigma}^2}\mathcal W^{e,i}_{\Sigma^2}\,.
\end{equation}
This entails that whenever one of the $n_i$'s is different from one,
we obtain a Hilbert space of dimension greater than one for the theory
on the boundary. From the point of view of the four dimensional
theory, these generalised Wilson lines are the operators measuring
background flux for the 1-form symmetries, so the fact that they do
not commute implies that we cannot choose Dirichlet boundary
conditions for all fluxes simultaneously, as argued originally in the
holographic context in \cite{Witten:1998wy}. In the specific case of
geometrically engineered four dimensional theories our field theory
discussion will reproduce the results obtained from geometry
previously in
\cite{Closset:2020scj,DelZotto:2020esg,Hosseini:2021ged}.

\section{Examples and consistency checks from $\mathcal N=2$ theories}\label{sec:examples}

In order to compute the Dirac pairing $\cQ$ we need to know the charge
lattice of the theory, meaning the electromagnetic charges of the
particles in the spectrum of the theory. In this paper, for
concreteness, we focus on examples arising in the context of 4d
$\mathcal N=2$ theories where we can easily extract this information
exploiting BPS quivers \cite{Alim:2011kw}. Here we are assuming that
the BPS spectrum faithfully reproduces the charge lattice, meaning
that we are assuming that in any charge sector populated by states we
can always find a BPS representative. By definition of BPS quiver the
Dirac pairing is captured by the quiver exchange matrix
\begin{equation}\label{eq:diraccher}
\mathcal Q_{ij} = \# (\text{ arrows } i \to j) - \# (\text{ arrows } j \to i) 
\end{equation}
where one works with an extended charge lattice with a number of generators that equals the number of electric, magnetic and flavour charges (where the latter are taken in the Cartan of the flavour symmetry $F$ of the theory). The generators of the charge lattice are in one to one correspondence with the nodes of the BPS quiver and all other stable BPS states have charges that can be expressed as linear combinations of the form $\gamma = \sum_{i=1}^{2r + f} M_i \gamma_i$ where $M_i \in \mathbb Z_{\geq 0}$ for all $i$ (particles) or $M_i \in \mathbb Z_{\leq 0}$ for all $i$ (antiparticles). In particular, we are granted that the $\gamma_i$ are in the spectrum and hence our screening argument is faithfully representing the 1-form symmetry.

\subsection{Pure $\cN=2$ SYM theories with algebra $\fg$}
We begin considering the center symmetries of pure SYM theories with
gauge algebra $\fg$.  The BPS quivers for all the ADE cases are
described in \cite{Cecotti:2010fi,Alim:2011kw} and those for the
non-simple laced cases where given in \cite{Cecotti:2012gh}. We
emphasize that in the latter case we have no IIB geometric engineering
construction for the BPS quiver, so the agreement that we will find
between the field theory expectation and the result from the analysis
in terms of the BPS quiver can be taken as evidence that our
discussion above remains valid for those cases for which the string
theory analysis is not available.

\begin{table}
\centering
\begin{tabular}{  c |  cl  }
\phantom{$\Big|$}$\fg$ & $\mathbb D^{(1)}$ \\
\hline
\hline
\phantom{$\Big|$} \multirow{1}{*}{$A_n$} 
         & $\bZ_{n+1}\oplus \bZ_{n+1}$ \\
\hline
\multirow{2}{*}{$D_n$} 
    & $\bZ_2\oplus\bZ_2 \oplus \bZ_2 \oplus \bZ_2$ & $\text{if }  \phantom{n}2 \mid n$\\
    & $\bZ_4\oplus \bZ_4$ & $\text{if }  \phantom{n}2 \nmid n$ \\
\hline
\phantom{$\Big|$}\multirow{1}{*}{$E_6$} 
         & $\bZ_3\oplus \bZ_3 $ \\
\hline
\phantom{$\Big|$}\multirow{1}{*}{$E_7\, ,\;B_n\, , \; C_n$} 
         & $\bZ_2\oplus \bZ_2$ \\
\hline
\phantom{$\Big|$}\multirow{1}{*}{$E_8\, ,\; G_2\, ,\; F_4$} 
         & $0$ \\
\end{tabular}
\caption{Defect groups for the pure $\cN=2$ SYM theories with algebra $\fg$.}\label{tab:rutta}
\label{table:pure SYM theories}
\end{table}

We find that for all these semi-simple Lie algebras, the Dirac pairing $\mathcal Q$ can be written in the block diagonal form and the corresponding defect groups are given in \ref{table:pure SYM theories}. This is clear from the structure of the Dirac pairing for such models, which is given by
\begin{equation}
\mathcal Q_{\fg} = \left( \begin{array}{c|c} C_\fg - C_\fg^t \phantom{\Big|} & \phantom{\Big|}  C_\fg^t   \\\hline \phantom{\Big|}   -C_\fg  & 0 \phantom{\Big|} \end{array}\right)
\end{equation}
where $C_\fg$ is the $r\times r$ Cartan matrix for the Lie algebra $\fg$. For example, the BPS quivers for gauge groups $SU(N+1)$ and $USp(2N)$ are
\begin{equation}\label{eq:QuiverEx}
\begin{gathered}
\xymatrix{&\bullet_1\ar[r]&\circ_2&\dots\ar[l]&\bullet_N\ar[l]\\
\\
&\circ_1\ar@<-0.2pc>[uu]\ar@<+0.2pc>[uu]&\bullet_2\ar@{<-}@<-0.2pc>[uu]\ar@{<-}@<+0.2pc>[uu]\ar[l]&\dots\ar@{<-}[l]&\circ_N\ar@<-0.2pc>[uu]\ar@<+0.2pc>[uu]\ar@{<-}[l]
}
\end{gathered}
\quad
\begin{gathered}
\xymatrix{&\bullet_1\ar[r]&\circ_2&\bullet_3\ar[l]\ar[r]&\dots&\bullet_N\ar[l]\\
\\
&\circ_1\ar[uur]\ar[uu]\ar@<+0.4pc>[uu]&\bullet_2\ar@{<-}@<-0.2pc>[uu]\ar@{<-}@<+0.2pc>[uu]\ar@<0.2pc>[l]\ar@<-0.2pc>[l]\ar[r]&\circ_3\ar@<-0.2pc>[uu]\ar@<+0.2pc>[uu]&\dots\ar@{<-}[l]&\circ_N\ar@<-0.2pc>[uu]\ar@<+0.2pc>[uu]\ar@{<-}[l]
}
\end{gathered}
\end{equation}
respectively. From these we obtain the block diagonal forms 
\begin{equation}
\widetilde{\mathcal Q}_{A_{N}}=\left(
\begin{array}{ccccc}
 0 & N+1 &  &  &  \\
 -(N+1) & 0 &  &  &  \\
  &  & 0 & 1 &  \\
  &  & -1 & 0 &  \\
   &  &  & & \ddots 
\end{array}
\right)
\qquad\qquad
\widetilde{\mathcal Q}_{C_N}=\left(
\begin{array}{ccccc}
 0 & 2 &  &  &  \\
 -2 & 0 &  &  &  \\
  &  & 0 & 1 &  \\
  &  & -1 & 0 &  \\
   &  &  & & \ddots 
\end{array}
\right)\, ,
\end{equation}
respectively. Thus, it is clear that, 
\begin{equation}
\mathbb D^{(1)}_{A_N} = (\bZ_{N+1})^{(1)}_e \oplus (\bZ_{N+1})^{(1)}_m \, ,\quad \mathbb D^{(1)}_{C_N} =(\bZ_{2})^{(1)}_e \oplus (\bZ_{2})^{(1)}_m \, .
\end{equation}
Moreover, the resulting Heisenberg algebra precisely reproduces the expected global forms for the corresponding Lie groups \cite{Garcia-Etxebarria:2019cnb}. The check for all other groups is carried in a similar fashion and we consistently recover all results in table \ref{tab:rutta}.


\subsection{Example of $\cN=2^*$ theories with algebra $\fg$}

In order to obtain the $\cN=2^*$ quivers for a simple gauge theory with gauge algebra of ADE type, one can start from the affine quiver $A(1,1) \square \widehat{G}$ and replace one of the Kronecker subquivers that correspond to a node with Dynkin weight 1 in the affine $\widehat{G}$ diagram with a single node $\ast$, giving a `specialization' of the corresponding quiver in the language of \cite{Cecotti:2012gh}. Graphically such operation corresponds to
\begin{equation}
\begin{gathered}
\xymatrix{\dots&\ar[r]\ar[l]\bullet&\dots\\
\\
\ar[r]\dots&\circ\ar@<0.2pc>[uu]\ar@<-0.2pc>[uu]&\ar[l]\dots}\end{gathered} \quad\longrightarrow\qquad \begin{gathered}\xymatrix{\dots&&\dots\\
&\ast\ar[ur]\ar[ul]&\\
\ar[ur]\dots&&\ar[ul]\dots}\end{gathered}
\end{equation}
Using this trick, for instance, the BPS quiver for $SU(N+1)$ $\cN=2^*$ is
\begin{equation}\label{eq:ADJSUN}
\begin{gathered}
\xymatrix{&\bullet_1\ar[dl]\ar[r]&\circ_2&\dots\ar[l]&\bullet_N\ar[l]\ar[dr]\\
\ast\ar[dr]&&&&&\ast\ar[dl]\\
&\circ_1\ar@<-0.2pc>[uu]\ar@<+0.2pc>[uu]&\bullet_2\ar@{<-}@<-0.2pc>[uu]\ar@{<-}@<+0.2pc>[uu]\ar[l]&\dots\ar@{<-}[l]&\circ_N\ar@<-0.2pc>[uu]\ar@<+0.2pc>[uu]\ar@{<-}[l]
}
\end{gathered}
\end{equation}
where the nodes $\ast$ appearing on the left and on the right of the above equation need to be identified. The resulting $\mathcal Q_{ij}$ can be read off straightforwardly from equation \eqref{eq:diraccher}. Similarly, the BPS quiver for $SO(2N)$ $\cN=2^*$ is
\smallskip
\begin{equation}
\begin{gathered}
\xymatrix{&\bullet_1&\ar@/_2.5pc/[dll]\circ_2\ar@{<-}[l]&\dots\ar[l]\circ_{N-2}&\bullet_N\ar[l]&\bullet_{N-1}\ar@/_1.5pc/[ll]\\
\ast\ar@/_2.5pc/[drr]&&&&&\\
&\circ_1\ar@<-0.2pc>[uu]\ar@<+0.2pc>[uu]&\bullet_2\ar@{<-}@<-0.2pc>[uu]\ar@{<-}@<+0.2pc>[uu]\ar[l]&\dots\ar@{<-}[l]\bullet_{N-2}\ar@{<-}@<-0.2pc>[uu]\ar@{<-}@<+0.2pc>[uu]&\circ_N\ar@<-0.2pc>[uu]\ar@<+0.2pc>[uu]\ar@{<-}[l]&\circ_{N-1}\ar@<-0.2pc>[uu]\ar@<+0.2pc>[uu]\ar@{<-}@/^1.5pc/[ll]\\
\\
}
\end{gathered}
\end{equation}
Also in these examples we obtain perfect agreement with the defect groups we expect from gauge theory. For instance starting from the quiver of $SU(N)$ $\mathcal N=2^*$ we obtain
\begin{equation}
\widetilde{\mathcal Q}_{SU(N) \, \mathcal N=2^*}=\left(
\begin{array}{ccccc}
 0 & N &  &  &  \\
 -N & 0 &  &  &  \\
  &  & 0 & 1 &  \\
  &  & -1 & 0 &  \\
   &  &  & & \ddots 
\end{array}
\right)
\end{equation}
as expected.

\subsection{Adding matter in various representations}
The BPS quiver for a gauge theory with gauge group $\mathfrak{g}$ and matter in an principal representation $R$ is easily obtained \cite{Alim:2011kw}. There is a one-to-one correspondence between principal representations and nodes of the Dynkin diagram of $\fg$. We can schematically summarize it as follows: the nodes of the Dynkin diagram are in one-to-one correspondence with the basis elements of the weight lattice, and principal representations are such that their highest weight $w(R_i)$ is $w(R_i) = \delta_{ij} \omega_j$, where $\omega_j$ is the weight basis. Graphically, the $i$-th node in the Dynkin diagram
\begin{equation}
\xymatrix{\circ\ar@{-}[r]&\circ\ar@{-}[r]&\cdots \ar@{-}[r]&\bullet_i\ar@{-}[r]&\cdots\ar@{-}[r]&\circ\ar@{-}[r]&\cdots}
\end{equation} 
correspond to the principal representation $R_i$. For the pure SYM BPS quiver of type $\fg$ there is a one-to-one correspondence between the nodes of the Dynkin diagram and full Kronecker subquivers. To obtain the quiver for SYM coupled to the $i$-th principal representation one adds a node to the BPS quiver connected to the $i$-th Kronecker subquiver as follows
\begin{equation}\label{eq:prince_matter}
\begin{gathered}
\xymatrix{\dots&\ar[r]\ar[l]\bullet&\dots\\
\\
\ar[r]\dots&\circ\ar@<0.2pc>[uu]\ar@<-0.2pc>[uu]&\ar[l]\dots}\end{gathered} \quad\longrightarrow\qquad \begin{gathered}
\xymatrix{\dots&\ar[r]\ar[l]\bullet\ar[dr]&\dots\\
&&\ast\ar[dl]\\
\ar[r]\dots&\circ\ar@<0.2pc>[uu]\ar@<-0.2pc>[uu]&\ar[l]\dots}
\end{gathered}
\end{equation}
Notice that this prescription is compatible with the prescription discussed in the previous section about $\cN=2^*$ theories whenever the adjoint is a principal representation (e.g. this is the case for $SO(2N)$ above).

\medskip

In general, if the extra matter corresponds to a tensor product of principal representations $R_i \otimes R_j$ the corresponding BPS quiver is obtained by connecting the extra node to the rest of the quiver with an oriented triangle for each of the corresponding Kroneckers. An example for this prescription is the BPS quiver for the adjoint of $\mathfrak{su}(N)$ in the previous section. The quiver in equation \eqref{eq:ADJSUN} is the BPS quiver for the representation
\begin{equation}
\mathbf{N} \otimes \overline{\mathbf{N}} = 1 \oplus \text{Adj}.
\end{equation}
and it corresponds to the tensor product of the fundamental $\xymatrix{\bullet_1 \ar@{-}[r]&\cdots}$ and the antifundamental $\xymatrix{\cdots&\bullet_N \ar@{-}[l]}$ representations of $SU(N)$.

\medskip

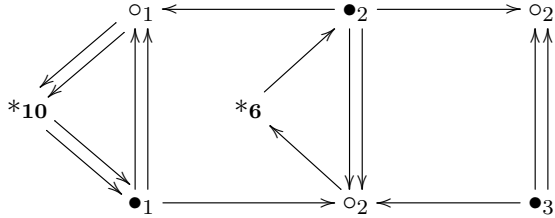
\begin{figure}
    \centering
    $$\xymatrix{
    & \circ_1 \ar@<0.2pc>[dl]\ar@<-0.2pc>[dl]&& \bullet_2 \ar[ll]\ar[rr]\ar@<0.2pc>[dd]\ar@<-0.2pc>[dd] && \circ_2\\
    \ast_\mathbf{10} \ar@<0.2pc>[dr]\ar@<-0.2pc>[dr] && \ast_\mathbf{6} \ar[ur] \\
    &\bullet_1\ar[rr] \ar@<0.2pc>[uu]\ar@<-0.2pc>[uu] && \circ_2 \ar[ul] &&  \bullet_3\ar[ll] \ar@<0.2pc>[uu]\ar@<-0.2pc>[uu] }$$
    \caption{The BPS quiver for $SU(4)$ with matter in the $\mathbf{6}\oplus \mathbf{10}$}
    \label{fig:SU4weird}
\end{figure}

It is amusing to check explicitly that the breaking of the center
symmetry by the $N$-ality of the corresponding representation is
respected, thus giving further consistency checks to our general
result. As an explicit example of how this works in practice let us
discuss here the case of the BPS quiver for a Lagrangian theory with
gauge group $SU(4)$ and matter in the direct sum of a symmetric
two-index representation of $SU(4)$ and an anti-symmetric two-index
representation. Since both the $\mathbf{6}$ and the $\mathbf{10}$ of
$SU(4)$ have quadrality 2, we expect the center symmetry of $SU(4)$ to
be broken down to $\mathbb Z_2$ in this example. The resulting quiver
is in Figure \ref{fig:SU4weird}. We indeed find a defect group
$\mathbb D^{(1)}$ given by two copies of $\bZ_2^{(1)}$ and a single
non-trivial BF coupling $n_1 = 2$, compatible with the field theory
expectations.

\subsection{Non-Lagrangian theories}

The results in this paper confirm field theoretically all the results
we obtained from studying the 4d $\mathcal N=2$ theories arising from
IIB on isolated hypersurface singularities. This follows from a simple
remark: in the IIB geometric engineering of both Lagrangian and
non-Lagrangian theories the Dirac pairing is captured by the
intersection form among the special Lagrangian vanishing 3-cycles of
the corresponding CY (see e.g. \cite{Shapere:1999xr}). In that context
the charge lattice of the theory is a sublattice of
$H_3(\mathcal X,\mathbb Z)$ given by stable collections of wrapped D3
branes. The quiver captures precisely this information, and the nodes
of the quiver give rise to a collection of 3-cycles that are always
stable, in regions of the moduli space that are compatible with that
quiver descriptions. The intersection pairing is therefore identified
with the Dirac pairing, and that is precisely the quantity which
enters in all the computations we carried out in our previous paper on
the subject and that determines the structure of the Heisenberg
algebra of FMS fluxes. For all these examples, therefore, our field
theoretical results and the ones obtained from geometric engineering
agree by construction. In particular, this confirms previous results
\cite{Closset:2020scj,DelZotto:2020esg,Hosseini:2021ged,Closset:2021lwy,Buican:2021xhs,Carta:2022spy,Bhardwaj:2021mzl}
about the Argyres-Douglas theories of type $(G,G')$ constructed by
Cecotti, Neitzke, and Vafa \cite{Cecotti:2010fi}, the various Arnol'd
SCFTs \cite{Cecotti:2011gu,DelZotto:2011an} and other theories
originating from singularity theory \cite{Xie:2015rpa}, as well as the
SCFTs of type $D_p(G)$ \cite{Cecotti:2012jx,Cecotti:2013lda}.

\medskip

We stress that our methods extend straightforwardly to all other theories with a known BPS quiver.

\medskip

As a further example of an application, we present in the rest of this section an analysis of the rank one theories with known BPS quivers, namely theories in the $\mathcal I_1$ series and in the $\mathcal I_4$ series in Table 1 of \cite{Argyres:2016xmc}. Therefore we obtain examples in all possible characteristic dimensions \cite{Cecotti:2021ouq}. The BPS quivers for these theories have been obtained in \cite{CDZ} --- see section 4 of \cite{Caorsi:2019vex} for a review (see also \cite{Alim:2011kw,Cecotti:2013lda,Cecotti:2013sza} for previous results on the topic). The resulting quivers have the form
\begin{equation}
    \begin{gathered}
        \xymatrix{\circ\ar[dd]|--{\underset{\quad}{\overset{\quad}{q_{\mathcal K}}}}\\
        &&\ast_1\ar[ull]|--{\,a_1\,}&\ast_2\ar[ulll]|--{\,a_2\,}&\cdots & \ast_f\ar[ulllll]|--{\,a_f\,}\\
        \bullet\ar[urr]|--{\,a_1\,}\ar[urrr]|--{\,a_2\,}\ar[urrrrr]|--{\,a_f\,}
        }
    \end{gathered}
\end{equation}
where:
\begin{itemize}
    \item  $f$ is the rank of the flavor symmetry of the rank 1 theory of interest;
    \item $q_\mathcal K$ is a positive integer denoting the multiplicity of the arrows $\circ \to \bullet$ determined as follows:
    \begin{equation}
        q_\mathcal K = \begin{cases}3 & \text{for } \mathcal K =II^*,III^*, IV^*\\ 2 & \text{for } \mathcal K = I^*_0 \\ 1 & \text{for } \mathcal K = II, III, IV \end{cases}
    \end{equation}
    \item $a_1,a_2,...,a_f$ are positive integers denoting the multiplicities of the arrows $\bullet \to \ast_i \to \circ$, determined from the decomposition of the Kodaira fiber
    \begin{equation}
        \mathcal K \to I_1,I_1,I_{(a_1)^2},\, I_{(a_2)^2},\,\dots ,\, I_{(a_f)^2}\,.
    \end{equation}
\end{itemize}
As an example for the $E_8$ Minhan-Nemeshansky theory \cite{Minahan:1996cj} we have $\mathcal K = II^*$ and $II^* \to (I_1)^{10}$, hence $q_{\mathcal K} = 3$, $f=8$ and $a_i = 1$ for all $i=1,...,8$. As another example, for the Argyres-Wittig theory with a flavor symmetry with Lie algebra $C_5$ we have $II^* \to (I_1)^{6},I_4$, hence $q_{\mathcal K} = 3$, $f=5$ and $a_1 = 2$ while $a_{2,3,4,5} = 1$.

\medskip

Exploiting the data of Table 1 of \cite{Argyres:2016xmc} it is straightforward to read off the corresponding BPS quivers for these theories, then by \eqref{eq:diraccher} the resulting BF levels follow by our method. The result we obtain is that for all these theories $n_1 = 1$, but for the case $\mathcal K = I_0^*$ with decomposition $I_0^* \to {I_1}^2,I_4$ which corresponds to $SU(2)$ $\mathcal N=2^*$. In this case we obtain $n_1=2$, as we already discussed above.

\medskip

The result we obtain for the global forms of these theories can be
also recovered exploiting the fact that these models arise as fixed
points of supersymmetry enhancing RG flows, starting from
$\mathcal N=1$ Lagrangians \cite{Gadde:2015xta}. Another interesting
class of susy enhancing RGs are those of Maruyoshi-Song type which
give results for the theories in the $\mathcal I_1$ series
\cite{Maruyoshi:2016tqk,Maruyoshi:2016aim,Agarwal:2018ejn}. For the
theory in the $\mathcal I_4$ series with global symmetry
$USp(4) \times U(1)$ as well as for the $E_6$ MN theory, UV
$\mathcal N=1$ theories have been obtained by brane bending and
deconfinement \cite{Etxebarria:2021lmq}. In all these examples we
reproduce easily the fact that the UV theory has a trivial defect
group, thus confirming our findings.

\section{Generalization to other dimensions}\label{sec:generaldims}
In this section we quickly comment about the generalization of the
argument above to some other theories in higher dimensions. In general
we expect the $D+1$ TFT action will contain terms of the following
form (with an additional factor of 2 in the self-dual case)
\begin{equation}
\begin{aligned}
\mathcal S^{TFT}_{D+1} \supseteq 2 \pi i \sum_{q=0}^{D}\mathcal Q^{\alpha,\beta}_q \int_{D+1} b^{(q+1)}_{\alpha} \wedge d b^{(D-q-1)}_{\beta}\,.
\end{aligned}
\end{equation}
The latter are relevant to probe more general global forms of QFTs in $D$-dimensions with different kinds of higher $q$-form symmetries corresponding to defects with non-trivial generalized Dirac strings. We stress that other couplings can be allowed, which in this paper we are omitting. In these cases the symmetry properties of the matrices $\mathcal Q^{\alpha,\beta}_q $ depend crucially on $D$ and on $q$. For instance for $D=6$ and $q=2$, we have a symmetric pairing that was explored in the context of defect groups and global structures of 6d (2,0) and (1,0) theories \cite{DelZotto:2015isa,GarciaEtxebarria:2019caf}. In what follows we quickly address the IR origin of the global structures for the cases of 6d (2,0) SCFTs and of 5d SCFTs.

\subsection{The case of 6d (2,0) theories}
In the case of 6d (2,0) theories we have an analogue of the Coulomb
branch, where the nonabelian string dynamics reduces to an abelian
one, the so-called \textit{tensor branch}. Along the tensor branch we
have a $(U(1)_e^{(2)})^r$ higher 2-form symmetry, which has 3-form
currents $J^{(3)} = h^{(3)}_i$ corresponding to the anti-self-dual
3-form curvatures $h_i^{(3)} = d b_i^{(2)}$ where $b_i^{(2)}$ are the
2-form fields in the 6d tensormultiples. We can couple the latter to
background fluxes $B^{(3)}_i$, which have 3-form background gauge
transformations analogous to the discussion we had for the Maxwell
theory. Also in this case, when we excite the (2,0) BPS strings, the
current conservation equation $d \ast J^{(3)} = 0$ is broken by the
presence of sources for the $h^{({3})}_i$ fluxes, represented by the
string charges.

The effect of such breaking is again detected by the inflow mechanism
which associates to the tensor branch a 7d BF like theory of the form
\begin{equation}\label{eq:7daction}
  \mathcal S_7 \supseteq {2\pi i \over 2} C^{\alpha \beta}_{\mathfrak g} \int_7 b^{(3)}_\alpha \wedge d b^{(3)}_\beta
\end{equation}
where $C^{\alpha \beta}_{\mathfrak g}$ is the Cartan matrix of the Lie
algebra of type $\mathfrak g$, which gives the BPS string Dirac
pairing for the (2,0) theory of type $\mathfrak g$ in 6d, again
accounting for the 't Hooft screening \cite{DelZotto:2015isa}.

There is a crucial difference between six dimensions and four
dimensions: in six dimensions the Dirac pairing in 6d is a symmetric
matrix, which is compatible with the symmetry properties of the 7-form
in the action \ref{eq:7daction}. For this reason, 6d strings can be
non-mutually local with respect to themselves. As examples one can
consider e.g. the rank one non Higgsable cluster theories, 6d SCFTs
with tensor branch of the form $\overset{\mathfrak{g}}{n}$. For these
models we obtain
\begin{equation}
    \mathcal S_7 \supseteq {2\pi i n \over 2} \int_7 b^{(3)} \wedge d b^{(3)} 
\end{equation}
This physical distinction comes together with a very important
mathematical distinction: while the theory of antisymmetric integral
bilinear forms relevant to the four dimensional case is very simple
(and in particular implies that a change of basis to the simple block
diagonal form~\eqref{eq:tilde-Q} always exists), the theory of
symmetric bilinear forms over the integers, relevant in the six
dimensional case, is significantly more complicated. For instance, one
can show that the Cartan matrix arising in the $\fsu(2)$ case cannot
be taken to diagonal form via an integral congruence
\cite{Conway1999}.

Interestingly, it is the diagonal form that appears in the holographic
result arising from the reduction of the Chern-Simons couplings of
M-theory \cite{Witten:1998wy,Aharony:1998qu}
\begin{equation}
    \mathcal S_7(A_{N-1}) \supseteq {2\pi i \over 2} N \int_7 c^{(3)} \wedge d c^{(3)} + ...
\end{equation}
We do not understand the physical meaning (if any) of this
discrepancy, we hope to return to it in future work.

\subsection{The case of 5d theories}
Another class of theories with interesting Coulomb phases is provided by the 5d SCFTs with non-trivial ranks. Along a 5d Coulomb branch we have an emergent $(U(1)^{(1)}_e \times U(1)^{(2)}_m)^r$ higher form symmetry. The latter is similarly broken to subgroups by the spectrum and, \textit{mutatis mutandis}, the same logic applies. The resulting BF-theory in this case has the form
\begin{equation}\label{eq:6daction}
\mathcal S_6 \supseteq {2\pi i} \mathcal Q^{\alpha \beta} \int_6 b^{(2)}_{e,\alpha} \wedge d b^{(3)}_{m,\beta}
\end{equation}
where $\mathcal Q^{\alpha \beta}$ is the Dirac pairing among the BPS
electric states and the 5d BPS monopole strings. This latter quantity
determines the structure of the Heisenberg algebras of non-commuting
fluxes via a field redefinition to its Smith normal form, thus
reproducing all the results obtained via M-theory in this context (see
e.g.
\cite{Albertini:2020mdx,Morrison:2020ool,Bhardwaj:2020phs,DelZotto:2022fnw,DelZotto:2022joo,Cvetic:2022imb}).\footnote{The
  Dirac pairing $\mathcal Q^{\alpha\beta}$ coincides with the intersection form
  between 2-cycles and 4-cycles relevant for the computation of the
  global forms from M-theory geometric engineering.} For this class of
theories, it is known there might be further terms in the bulk TFT
\cite{Apruzzi:2021nmk}: it should be possible to recover the latter in
terms of the infrared as well, however this would go beyond the modest
scope of this short note.

\medskip

As a consistency check of the above formula, one can consider the 4d KK theory of the corresponding SCFT, obtained from circle reduction. These systems have 5d BPS quivers \cite{Closset:2019juk}. The latter can be exploited in a way analogous to the one we discussed above to capture the global structure of the 4d KK theory so obtained. See reference  \cite{DelZotto:2022fnw} for an application of this idea in the context of 5d orbifold SCFTs discussed e.g. in \cite{Tian:2021cif}.

\section{Conclusions}\label{sec:conclusions}

In this paper we have begun exploring a mechanism to recover the
global structure of a given QFT from an infrared phase which is under
perturbative control. Our main result is to recover, from Coulomb-like
phases, the Heisenberg algebra of non-commuting fluxes that was found
in the geometric engineering analysis in purely field theoretical
terms.

\medskip

An interesting question that we leave for future analysis is to
characterize the full structure of the symmetry TFT from an IR
perspective. We expect this to be possible by a suitable extension of
the 't Hooft anomaly matching argument: while the theory
on the boundary flows, the symmetry TFT in the bulk must match along
the flow. In this short note, we recovered the term responsible for
the possible choices of global structures, but we stress that, for
instance, we expect mixing terms between the various higher form
fields in the symmetry TFT. The latter are not captured by the
argument presented here, but are known to arise from a geometric
engineering perspective \cite{Apruzzi:2021nmk,Apruzzi:2022dlm}.

\medskip

In this context a direction that we find particularly interesting is
the question of recovering higher group structures or more general
non-invertible symmetries from the IR. For two-groups, evidence that
this is indeed possible in some cases was given in the context of
little string theories in the papers
\cite{Cordova:2018cvg,Cordova:2020tij,DelZotto:2020sop}. There it was
shown that higher group structure constants are related to specific
terms in the anomaly polynomial of the corresponding little
strings. Based on that analysis we conjecture the anomaly theories on
the worldvolumes of the various BPS degrees of freedom of the boundary
QFTs of interest must know about these finer details of the symmetry
TFT. A similar effect was recently exploited to unravel certain
non-invertible symmetries in \cite{Kaidi:2021xfk,Choi:2021kmx}.

\acknowledgments

We thank P.~Argyres and M.~Martone for illuminating discussions. This project has received funding from the European Research Council
(ERC) under the European Union's Horizon 2020 research and innovation
programme (grant agreement No. 851931). I.G.-E. is partially supported
by STFC consolidated grant ST/T000708/1. We also both acknowledge
support from the Simons Foundation via the Simons Collaboration on
Global Categorical Symmetries.

\bibliographystyle{JHEP}
\bibliography{refs}

\end{document}